\documentclass[aps,prl,twocolumn,showpacs,preprintnumbers,amsmath,amssymb,superscriptaddress]{revtex4}

\usepackage{graphicx}% Include figure files
\usepackage{dcolumn}% Align table columns on decimal point
\usepackage{bm}% bold math
\usepackage{xspace}

\begin{document}

\newcommand{\superk}    {Super-Kamiokande\xspace}       
\newcommand{\nue}       {$\nu_{e}$\xspace}
\newcommand{\numu}      {$\nu_{\mu}$\xspace}
\newcommand{\nutau}     {$\nu_{\tau}$\xspace}
\newcommand{\nusterile} {$\nu_{sterile}$\xspace}
\newcommand{\mutau}     {$\nu_\mu \rightarrow \nu_{\tau}$\xspace}
\newcommand{\musterile} {$\nu_\mu \rightarrow \nu_{sterile}$\xspace}
\newcommand{\dms}       {$\Delta m^2$\xspace}
\newcommand{\sstt}      {$\sin^2 2 \theta$\xspace}
\newcommand{\Rnqe}      {$R_{\rm nqe}$\xspace}

\title{Search for coherent charged pion production in neutrino-carbon interactions}

\newcommand{\BCN}{\affiliation{Institut de Fisica d'Altes Energies, Universitat Autonoma de Barcelona, E-08193 Bellaterra (Barcelona), Spain}}
\newcommand{\BU}{\affiliation{Department of Physics, Boston University, Boston, Massachusetts 02215, USA}}
\newcommand{\UBC}{\affiliation{Department of Physics \& Astronomy, University of British Columbia, Vancouver, British Columbia V6T 1Z1, Canada}}
\newcommand{\UCI}{\affiliation{Department of Physics and Astronomy, University of California, Irvine, Irvine, California 92697-4575, USA}}
\newcommand{\SACLAY}{\affiliation{DAPNIA, CEA Saclay, 91191 Gif-sur-Yvette Cedex, France}}
\newcommand{\CNU}{\affiliation{Department of Physics, Chonnam National University, Kwangju 500-757, Korea}}
\newcommand{\DU}{\affiliation{Department of Physics, Dongshin University, Naju 520-714, Korea}}
\newcommand{\DUKE}{\affiliation{Department of Physics, Duke University, Durham, North Carolina 27708, USA}}
\newcommand{\GENEVA}{\affiliation{DPNC, Section de Physique, University of Geneva, CH1211, Geneva 4, Switzerland}}
\newcommand{\UH}{\affiliation{Department of Physics and Astronomy, University of Hawaii, Honolulu, Hawaii 96822, USA}}
\newcommand{\KEK}{\affiliation{High Energy Accelerator Research Organization(KEK), Tsukuba, Ibaraki 305-0801, Japan}}
\newcommand{\HIR}{\affiliation{Graduate School of Advanced Sciences of Matter, Hiroshima University, Higashi-Hiroshima, Hiroshima 739-8530, Japan}}
\newcommand{\INR}{\affiliation{Institute for Nuclear Research, Moscow 117312, Russia}}
\newcommand{\KOBE}{\affiliation{Kobe University, Kobe, Hyogo 657-8501, Japan}}
\newcommand{\KOR}{\affiliation{Department of Physics, Korea University, Seoul 136-701, Korea}}
\newcommand{\KYO}{\affiliation{Department of Physics, Kyoto University, Kyoto 606-8502, Japan}}
\newcommand{\LSU}{\affiliation{Department of Physics and Astronomy, Louisiana State University, Baton Rouge, Louisiana 70803-4001, USA}}
\newcommand{\MIT}{\affiliation{Department of Physics, Massachusetts Institute of Technology, Cambridge, Massachusetts 02139, USA}}
\newcommand{\MIYAGI}{\affiliation{Department of Physics, Miyagi University of Education, Sendai 980-0845, Japan}}
\newcommand{\NIIGATA}{\affiliation{Department of Physics, Niigata University, Niigata, Niigata 950-2181, Japan}}
\newcommand{\OKAYAMA}{\affiliation{Department of Physics, Okayama University, Okayama, Okayama 700-8530, Japan}}
\newcommand{\OSAKA}{\affiliation{Department of Physics, Osaka University, Toyonaka, Osaka 560-0043, Japan}}
\newcommand{\ROME}{\affiliation{University of Rome La Sapienza and INFN, I-000185 Rome, Italy}}
\newcommand{\SNU}{\affiliation{Department of Physics, Seoul National University, Seoul 151-747, Korea}}
\newcommand{\SOLTAN}{\affiliation{A.~Soltan Institute for Nuclear Studies, 00-681 Warsaw, Poland}}
\newcommand{\TOHOKU}{\affiliation{Research Center for Neutrino Science, Tohoku University, Sendai, Miyagi 980-8578, Japan}}
\newcommand{\SB}{\affiliation{Department of Physics and Astronomy, State University of New York, Stony Brook, New York 11794-3800, USA}}
\newcommand{\TUS}{\affiliation{Department of Physics, Tokyo University of Science, Noda, Chiba 278-0022, Japan}}
\newcommand{\KAM}{\affiliation{Kamioka Observatory, Institute for Cosmic Ray Research, University of Tokyo, Kamioka, Gifu 506-1205, Japan}}
\newcommand{\RCCN}{\affiliation{Research Center for Cosmic Neutrinos, Institute for Cosmic Ray Research, University of Tokyo, Kashiwa, Chiba 277-8582, Japan}}
\newcommand{\TRIUMF}{\affiliation{TRIUMF, Vancouver, British Columbia V6T 2A3, Canada}}
\newcommand{\VAL}{\affiliation{Instituto de F\'{i}sica Corpuscular, E-46071 Valencia, Spain}}
\newcommand{\UW}{\affiliation{Department of Physics, University of Washington, Seattle, Washington 98195-1560, USA}}
\newcommand{\WARSAW}{\affiliation{Institute of Experimental Physics, Warsaw University, 00-681 Warsaw, Poland}}

% to make affiliation numbers in order
\BCN
\BU
\UBC
\UCI
\SACLAY
\CNU
\DU
\DUKE
\GENEVA
\UH
\KEK
\HIR
\INR
\KOBE
\KOR
\KYO
\LSU
\MIT
\MIYAGI
\NIIGATA
\OKAYAMA
\OSAKA
\ROME
\SNU
\SOLTAN
\TOHOKU
\SB
\TUS
\KAM
\RCCN
\TRIUMF
\VAL
\UW
\WARSAW

\author{M.~Hasegawa}\KYO 
\author{E.~Aliu}\BCN                
\author{S.~Andringa}\BCN 
\author{S.~Aoki}\KOBE 
\author{J.~Argyriades}\SACLAY 
\author{K.~Asakura}\KOBE 
\author{R.~Ashie}\KAM 
\author{H.~Berns}\UW 
\author{H.~Bhang}\SNU 
\author{A.~Blondel}\GENEVA 
% ...
\collaboration{The K2K Collaboration}\noaffiliation

\date{\today}%  It is always \today, today,

\begin{abstract}
%\input{abstract}

%%%%%%%%%%%%%%%%%%%%%%%%%%%%%%%%%%%%%%%%%%%%%%%%%%%%%%%%%%%%%%%%%%%%%%%
%\chapter{abstract}
%%%%%%%%%%%%%%%%%%%%%%%%%%%%%%%%%%%%%%%%%%%%%%%%%%%%%%%%%%%%%%%%%%%%%%%

 We report the result from a search for charged-current coherent pion 
production induced by muon neutrinos with a mean energy of 1.3~GeV. 
 The data are collected with a fully active scintillator detector 
in the K2K long-baseline neutrino oscillation experiment.
 No evidence for coherent pion production is observed and an 
upper limit of $0.60 \times 10^{-2}$ is set on the 
cross section ratio of coherent pion production to the 
total charged-current interaction at 90\% confidence level. 
 This is the first experimental limit for coherent charged pion
production in the energy region of a few GeV.

\end{abstract}

\maketitle

%%%%%%%%%%%%%%%%%%%%%%%%%%%%%%%%%%%%%%%%%%%%%%%%%%%%%%%%%%%%%%%%%%%%%
%\section{Introduction}
%%%%%%%%%%%%%%%%%%%%%%%%%%%%%%%%%%%%%%%%%%%%%%%%%%%%%%%%%%%%%%%%%%%%%

 The charged-current (CC) coherent pion production in neutrino-nucleus
scattering, $\nu_{\mu} + A \to \mu^{-} + \pi^{+} + A$, is 
a process in which the neutrino scatters coherently off 
the entire nucleus with a small energy transfer.
 Such a process has been measured in a number of
experiments~\cite{Vilain:1993pf, Allport:1989pf,
Grabosch:1986pf, Willocq:1993pf}, 
providing a test of the partially conserved axial-vector 
current~(PCAC) hypothesis~\cite{Adler:1964}. 
 The existing data agree with the Rein and Sehgal model~\cite{Rein:1982pf}
based on the PCAC hypothesis for neutrino energies from 7 to 100
GeV, while there exists no measurement at lower energies.

 The recent discovery of neutrino oscillations has renewed interest
in neutrino-nucleus interactions in the sub- to few GeV region.
 The KEK to Kamioka~(K2K) long-baseline 
neutrino oscillation experiment has reported~\cite{Aliu:2004} 
a significant deficit in the forward scattering events,
which limits the prediction accuracy of the neutrino energy spectrum at
the far detector.
 CC coherent pion production is one of the candidate interactions 
responsible for  this deficit
and its study is necessary to improve the accuracy of the current 
and future atmospheric/accelerator-based neutrino oscillation
experiments, which are expected to achieve much improved statistical 
precision using interactions of neutrinos in the same energy region as K2K. 

 This letter presents the result from a search for CC
coherent pion production by neutrinos in the K2K experiment.
 We compare our result specifically with the Rein and Sehgal 
model~\cite{Rein:1982pf} because it is the only model that provides 
the kinematics of pions and is commonly used in neutrino 
oscillation experiments.

%%%%%%%%%%%%%%%%%%%%%%%%%%%%%%%%%%%%%%%%%%%%%%%%%%%%%%%%%%%%%%%%%%%%%%
%\section{K2K experiment}
%%%%%%%%%%%%%%%%%%%%%%%%%%%%%%%%%%%%%%%%%%%%%%%%%%%%%%%%%%%%%%%%%%%%%%

 In the K2K experiment, protons are extracted 
from the KEK 12~GeV proton synchrotron and hit an aluminum target.
 Positively charged secondary particles, mainly pions, are focused by a
magnetic horn system and decay to produce
 an almost pure~(98\%) $\nu_{\mu}$ beam with a mean energy of 
1.3~GeV~\cite{Ahn:2001cq}.
 The neutrino beam energy spectrum and spatial profile are measured
using a set of near neutrino detectors located 300~m downstream from 
the proton target.
 The estimated absolute flux has a large uncertainty due to
difficulties in the absolute estimation of the primary proton beam intensity, 
the proton targeting efficiency, and hadron production cross sections. 
 Therefore, the ratio of the CC coherent pion to the total CC cross
section is measured, rather than the absolute CC coherent pion cross
section.
 The data used for this analysis were collected with one of the near 
detectors, the fully active scintillator detector~(SciBar), 
from October 2003 to February 2004, 
corresponding to $1.7 \times 10^{19}$ protons on target~(POT).

 The SciBar detector~\cite{Nitta:2004nt} consists of 14,848 extruded
plastic scintillator strips read out by wavelength-shifting fibers 
and multi-anode photomultipliers. 
 The scintillator also acts as the neutrino interaction target; 
it is a fully active detector and has high efficiency for low 
momentum particles.  
 Scintillator strips with dimensions of $1.3\times2.5\times300~\mathrm{cm}^{3}$ are 
arranged in 64 layers. 
 Each layer consists of two planes to measure horizontal and vertical
position. 
 The total size of the detector is $3.0\times3.0\times1.7~\mathrm{m}^{3}$,
while an inner volume of $2.6\times2.6\times1.35~\mathrm{m}^{3}$~(9.38 tons) 
is used as the fiducial volume to reject incoming particles and obtain a
flat efficiency for CC interactions. 
 The minimum reconstructible track length is 8~cm.
 A track finding efficiency of more than 99\% is achieved 
for single tracks with a length of more than 10~cm.
 The track finding efficiency for a second, shorter track is 
lower than that for single tracks due to overlap with the first track. 
 This efficiency increases with the length of the second track and reaches 
90\% at a track length of 30~cm.

%%%%%%%%%%%%%%%%%%%%%%%%%%%%%%%%%%%%%%%%%%%%%%%%%%%%%%%%%%%                
%\section{MC simulation}
%%%%%%%%%%%%%%%%%%%%%%%%%%%%%%%%%%%%%%%%%%%%%%%%%%%%%%%%%%%                                                          
 The NEUT Monte Carlo~(MC) simulation program library~\cite{hayato:2002} 
is used to simulate neutrino-nucleus interactions. 
The CC coherent pion production is incorporated in the simulation 
based on the Rein and Sehgal model~\cite{Rein:1982pf},
 which predicts the cross 
section averaged over the K2K neutrino energy spectrum 
of $2.85\times 10^{-40} \mathrm{cm}^{2}$/nucleon for carbon.
 The Llewellyn Smith model~\cite{Llewellyn:1972} and the Rein and
Sehgal model~\cite{Rein:1981} are employed for quasi-elastic~(QE)
scattering~($\nu_{\mu} + n \to \mu^{-} + p$) and 
CC single pion (1$\pi$) production~($\nu_{\mu} + N \to \mu + N + \pi$),
where \textit{N} is a nucleon, respectively.
 The axial vector mass of the nucleon form factor is 
set to be 1.1~$\mathrm{GeV/{\it c}}^{2}$ for both QE and CC1$\pi$ 
interactions~\cite{Bernard:2002}.
 For deep inelastic scattering (DIS), we use GRV94 nucleon structure
functions~\cite{GRV:1995} with a correction by Bodek and
Yang~\cite{Bodek:2002}. 
 Nuclear effects are taken into account;
 for the pions originating from neutrino interactions,
absorption, elastic scattering, and charge exchange
inside the target nucleus are simulated.
 Pion cross sections are calculated using
the model by Salcedo et al.~\cite{Salcedo:1988}, 
which agrees well with past experimental data~\cite{Ingram:1983}.
 Pion interactions outside the target nucleus are simulated 
based on other experimental data~\cite{Carroll:1976}.

%%%%%%%%%%%%%%%%%%%%%%%%%%%%%%%%%%%%%%%%%%%%%%%%%%%%%%%%%%%%%%%%%%%%%
%
%%%%%%%%%%%%%%%%%%%%%%%%%%%%%%%%%%%%%%%%%%%%%%%%%%%%%%%%%%%%%%%%%%%%%

 For the present analysis, the experimental signatures of CC coherent 
pion production are the existence of exactly two tracks, both consistent 
with minimum ionizing particles, and small momentum transfer defined 
as $q^{2} \equiv (P_{\mu}-P_{\nu})^{2}$, where $P_{\mu}$ and $P_{\nu}$ 
are the four momenta of the muon and the neutrino, respectively.
 According to the MC simulation, the dominant background 
is the CC1$\pi$ production, where the proton is below threshold or 
the neutron is invisible. 

%%%%%%%%%%%%%%%%%%%%%%%%%%%%%%%%%%%%%%%%%%%%%%%%%%%%%%%%%%%%%%%%%%
%\section{Event selection 1}
%%%%%%%%%%%%%%%%%%%%%%%%%%%%%%%%%%%%%%%%%%%%%%%%%%%%%%%%%%%%%%%%%%

 Charged current~(CC) candidate events are selected by requiring that at least one
reconstructed track starting in the fiducial volume is
matched with a track or hits in the muon range
detector~(MRD)~\cite{Ishii:2001sj} located just behind
SciBar~(SciBar-MRD sample).  
 This criterion imposes a threshold for muon momentum~($p_{\mu}$) of 450~MeV/c. 
 According to the MC simulation, 98\% of the events selected by this 
requirement are CC induced events, and the rest are neutral current~(NC) 
interactions accompanied by a charged pion or proton which penetrates 
into the MRD. 
 The contribution from $\nu_{e}$ is negligible~($<0.4\%$).
 The momentum of the muon is reconstructed from its range through SciBar
and MRD.
 The resolutions for $p_{\mu}$ and the angle with respect to
the neutrino beam direction~($\theta_{\mu}$) are determined to be 
80~MeV/c and 1.6~degrees, respectively.
%%%%%%%%%%%%%%%%%%%%%%%%%%%%%%%%%%%%%%%%%%%%%%%%%%%%%%%%%%%%%%%%%%%%%
%%%%%%%%%%%%%%%%%%%%%%%%%%%%%%%%%%%%%%%%%%%%%%%%%%%%%%%%%%%%%%%%%%%%%

 From the SciBar-MRD sample, events with two reconstructed tracks
are selected.
The QE candidate events are rejected 
by using kinematic information~\cite{Aliu:2004}. 
 Events in which the shorter track is identified as proton-like
based on \textit{dE/dx} information~(\textit{non-QE-proton} sample) 
are also rejected to select the \textit{non-QE-pion} sample, 
which includes the signal candidates.
 The particle identification capability is verified using 
cosmic ray muons and the shorter tracks in the QE sample, where the
latter provides a proton sample with more than 90\%.
 The probability to mis-identify a muon track as
proton-like is 1.7\% with a corresponding proton selection efficiency 
of 90\%.

%%%%%%%%%%%%%%%%%%%%%%%%%%%%%%%%%%%%%%%%%%%%%%%%%%%%%%%%%%%%%%%%%%%%%%%
%\section{Coherent pion yield}
%%%%%%%%%%%%%%%%%%%%%%%%%%%%%%%%%%%%%%%%%%%%%%%%%%%%%%%%%%%%%%%%%%%%%%%

\newcommand{\evtx}{\ensuremath{E_\mathrm{vtx}}}

\begin{figure} [htpb]
 \begin{center}
  \resizebox{8.0cm}{!}{
   \includegraphics{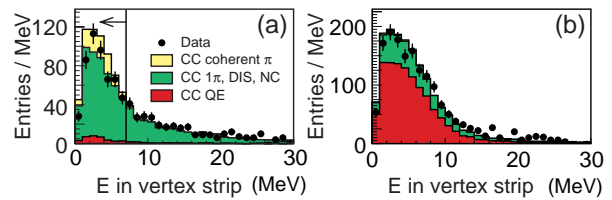}}
   \caption{Distribution of $\evtx$ 
   for (a) non-QE-pion sample and (b) QE sample.
   Black circles: observed data, Histograms: MC
  expectation with breakdown of interaction modes.
   The statistical $\chi^{2}$/DOF in the selected region 
   of (a), indicated by a vertical line, is 30.1/7~(9.8/7) 
   with~(without) CC coherent pion production.}
   \label{fig:vac}
 \end{center}
\end{figure}

The CC coherent pion candidates are extracted from the non-QE-pion
sample.
The background events are suppressed by requiring that the pion-like track
goes forward. 
 Even if the additional particles in the background process are not reconstructed 
as tracks, they can be detected as a large energy deposit 
or additional hits around the vertex.
 Figure~\ref{fig:vac}(a) shows a distribution of energy deposited in the vertex
strip ($\evtx$) for the non-QE-pion sample. 
 The MC prediction for $\evtx$ is verified with the QE sample, which has
no contribution from non-visible particles, as shown in Fig.~1(b).  
 We require the events to have $\evtx$ less than 7~MeV and 
no additional hits around the vertex strip.

 The value of $q^{2}$ reconstructed from $p_{\mu}$ and $\theta_{\mu}$
under the assumption of QE interaction is denoted
$q^{2}_{\mathrm{rec}}$, and is calculated using            
\begin{eqnarray*}                                                                  
  p_{\nu} &=& \frac{1}{2}\frac{(M_{p}^{2}-m_{\mu}^{2})+2E_{\mu}(M_{n}-V)-(M_{n}-V)^{2}}{-E_{\mu\
}+(M_{n}-V)+p_{\mu}\cos\theta_{\mu}}                                                  
\end{eqnarray*}                                                                    
 where $M_{p(n)}$ is the proton~(neutron) mass, $m_{\mu}$ is the muon
mass and \textit{V} is the nuclear potential set to 27~MeV.                           
 The $q^{2}_{\mathrm{rec}}$ for coherent pion production events, which
is expected to be very small due to the small scattering angle for
muons, is shifted from the true $q^{2}$ by 0.008 $\mathrm{(GeV/{\it c})}^{2}$
with a resolution of 0.014 $\mathrm{(GeV/{\it c})}^{2}$.
 Events are required to have a reconstructed $q^{2}$ of less than 
0.10~$\mathrm{(GeV/{\it c})}^{2}$.

\begin{figure} [tbph]
 \begin{center}
  \resizebox{8.5cm}{!}{
   \includegraphics{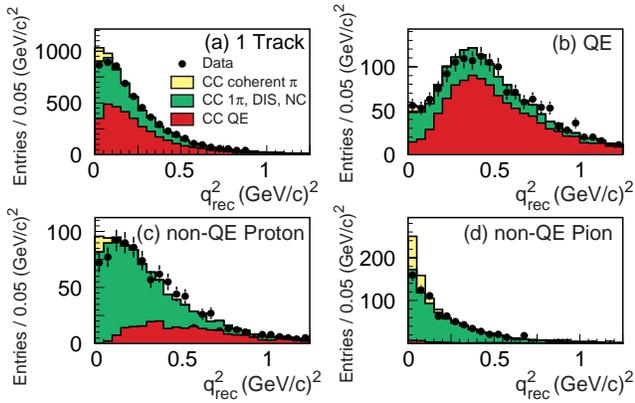}}
   \caption{The $q^{2}_{\mathrm{rec}}$ distributions for the (a) 1track, (b)
   QE, (c) non-QE-proton, and (d) non-QE-pion samples. 
   The statistical $\chi^{2}$/DOF in the region $q^{2}_{\mathrm{rec}} <
   0.10~(\mathrm{GeV/c})^{2}$ of (c) and (d) are 7.2/2~(2.7/2) and 
   32.3/2~(1.2/2) with~(without) CC coherent pion production.}
   \label{q2after}
 \end{center}
\end{figure}

 The background contamination in the final sample is estimated by the MC
simulation. 
 In order to constrain the uncertainties,
the $q^{2}_{\mathrm{rec}}$ distributions of the data
in the region $q^{2}_{\mathrm{rec}} > 0.10 \mathrm{(GeV/{\it c})}^{2}$
are fitted with MC expectations.
 The one track sample is used as well as two-track QE, non-QE-proton and
 non-QE-pion samples, and these four samples are fitted simultaneously.
 In the fit, the non-QE to QE relative cross section ratio, 
the magnitude of the nuclear effects and the momentum scale for muons 
are treated as free parameters.  
 Figure~\ref{q2after} shows the $q^{2}_{\mathrm{rec}}$ distributions
of the data with the MC simulation after the fitting.
 The $\chi^{2}$ value in the regions with
$q^{2}_{\mathrm{rec}}> 0.10 \mathrm{(GeV/{\it c})}^{2}$  at the best fit 
is 73.2 for 82 degrees of freedom (DOF).

\begin{figure} [htpb]
 \begin{center}
  \resizebox{6.0cm}{!}{
   \includegraphics{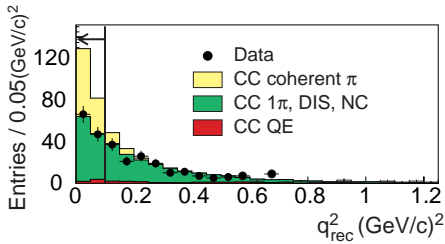}}
   \caption{The reconstructed $q^{2}$ distribution in the final
   sample. }
   \label{fig:q2final}
 \end{center}
\end{figure}

\begin{figure} [htpb]
 \begin{center}
 \resizebox{8.5cm}{!}{
   \includegraphics[trim=0.0cm 0.0cm 0.0cm
   0.0cm,clip]{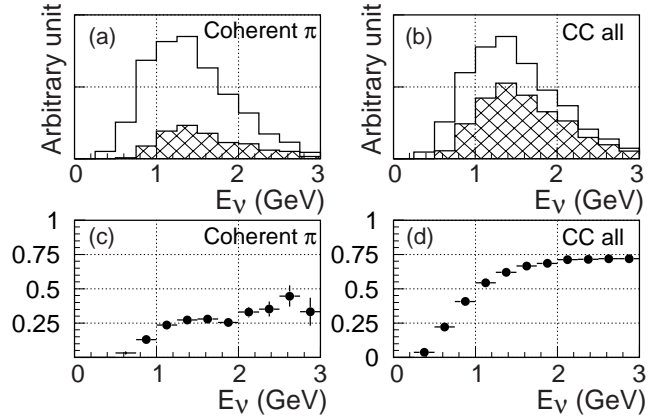}}
   \caption{Top: The neutrino energy spectra for (a) the coherent pion
   and (b) total CC events.
   The hatched histograms show the selected events. Bottom: The efficiencies
   as a function of neutrino energy for (c) the coherent pion and (d)
   total CC events. All of them are estimated by the MC simulation.}
   \label{fig:effcurve}
 \end{center}
\end{figure}

 Figure \ref{fig:q2final} shows the $q^{2}_{\mathrm{rec}}$ 
distribution for the final CC coherent pion sample. 
 The number of events in each selection step is
summarized in Table~\ref{table:eventsummary} together with the
signal efficiency and purity.
 In the signal region, 113 coherent pion candidates are found.
 The neutrino energy spectra for coherent pion events and the efficiency 
as a function of neutrino energy, estimated using the MC simulation,
are shown in Fig.~\ref{fig:effcurve}(a) and \ref{fig:effcurve}(c),
respectively.
 The total efficiency is 21.1\%.
 The expected number of background events in the signal region is 111.4.    
 After subtracting the background and correcting for the efficiency, 
the number of coherent pion events is measured to be 
$7.64\pm 50.40~(\mathrm{stat.})$, while 470 events are
expected from the MC simulation. 
 Hence, no evidence of coherent pion production is 
found in the present data set.

\begin{table} [htp]
\begin{center}
 \begin{tabular}{lccr}
\hline
\hline
            & Data & Efficiency  & Purity  \\
            &      &    (\%)     & (\%)    \\
\hline
 SciBar-MRD             & 10049& 77.9 & 3.6 \\
 Two track              & 3396 & 35.5 & 5.1 \\
 Non-QE pion            &  843 & 27.7 & 14.8 \\
 Second track direction & 773  & 27.3 & 15.8 \\
 No activity around the vertex & 297 & 23.9 & 28.2 \\
 $q^{2}_{\mathrm{rec}} \le 0.10 \mathrm{(GeV/{\it c})}^{2}$ & 113 & 21.1 &
  47.1 \\ 
 \hline
 \hline
 \end{tabular}  
 \caption{The number of events, the MC efficiency and purity of coherent
  pion events after each selection step.
}
 \label{table:eventsummary}
 \end{center}
 \end{table}

%%%%%%%%%%%%%%%%%%%%%%%%%%%%%%%%%%%%%%%%%%%%%%%%%%%%%%%%%%%%%%%%%%%%%%%
%\section{Total CC}
%%%%%%%%%%%%%%%%%%%%%%%%%%%%%%%%%%%%%%%%%%%%%%%%%%%%%%%%%%%%%%%%%%%%%%%

 The total number of CC interactions is estimated by using the SciBar-MRD
sample.
 As shown in Table I, 10049 events fall into this category.
 Based on the MC simulation, the selection efficiency and purity for 
CC interactions in the sample are estimated to be 56.9\% and 98.0\%, 
respectively.
 The expected neutrino energy spectra and 
the energy dependence of the selection efficiency for CC events are shown in
Fig.~\ref{fig:effcurve}(b) and \ref{fig:effcurve}(d), respectively.
The total number of CC events
is obtained to be $(1.73\pm0.02(\mathrm{stat.}))\times10^{4}$. 
 We derive the cross section ratio of CC coherent
pion production to the total CC interaction to be $(0.04\pm
0.29(\mathrm{stat.}))\times 10^{-2}$.

%%%%%%%%%%%%%%%%%%%%%%%%%%%%%%%%%%%%%%%%%%%%%%%%%%%%%%%%%%%%%%%%%%%%%%%
%\section{Systematic Error}
%%%%%%%%%%%%%%%%%%%%%%%%%%%%%%%%%%%%%%%%%%%%%%%%%%%%%%%%%%%%%%%%%%%%%%%

 Systematic uncertainties for the cross section ratio 
are summarized in Table~\ref{syserror}. 
 The major contributions come from uncertainties of nuclear 
effects and the neutrino interaction models. 
 The uncertainty due to nuclear effects is estimated
by varying the cross sections of pion absorption and elastic scattering 
by $\pm$30\% based on the accuracy of the reference data~\cite{Ingram:1983}.
 The uncertainties in QE and CC1$\pi$ interactions are estimated by
changing the axial vector mass by $\pm$ 0.10 $\mathrm{GeV/{\it c}}^{2}$~\cite{Bernard:2002}. 
 For DIS, the effect of the Bodek and Yang
correction is evaluated by changing the amount of correction by $\pm$30\%. 
 The $q^{2}_{\mathrm{rec}}$ distribution of the non-QE-proton sample (Fig.~\ref{q2after}(c))
indicates an additional deficit of background events in the region
$q^{2}_{\mathrm{rec}}< 0.10~(\mathrm{GeV/{\it c}})^{2}$.
 CC1$\pi$ interaction dominates events in this region;
its cross section has significant uncertainty due to nuclear effects.
 We estimate the amount of possible deficit in the same manner as 
described in~\cite{Aliu:2004} with the one track, 
QE and non-QE-proton samples. 
 We find that a 20\% suppression of CC1$\pi$ events for
$q^{2}_{\mathrm{true}} < 0.10~(\mathrm{GeV/{\it c}})^{2}$ is allowed, 
which varies the cross section ratio by  $+0.14\times 10^{-2}$.
This variation is conservatively treated as a systematic uncertainty.
 We also consider the uncertainties of the event selection, 
where the dominant error comes from track counting,
detector response such as scintillator quenching, and 
neutrino energy spectrum shape. 
The total systematic uncertainty on the cross section ratio amounts to 
$+0.32$/$-0.35 \times 10^{-2}$.

\begin{table} [htp]
\begin{center}
 \begin{tabular}{lcc}
 \hline
 \hline
 Error source  & \multicolumn{2}{c}{Uncertainty of $\sigma$ ratio ($\times 10^{-2}$)} \\
 \hline
 Nuclear effects    &~~~~~ +0.23 & $-0.24$ \\
 Interaction model &~~~~~ +0.10 & $-0.09$ \\
 CC1$\pi$ suppression &~~~~~ +0.14 & --- \\  
 Event selection   &~~~~~ +0.11 & $-0.17$ \\
 Detector response &~~~~~ +0.09 & $-0.16$ \\
 Energy spectrum   &~~~~~ +0.03 & $-0.03$ \\   
 \hline
 Total &~~~~~ +0.32 & $-0.35$ \\
 \hline
 \hline
 \end{tabular}
 \caption{The summary of systematic uncertainties in the 
 (CC coherent pion)/(total CC interaction) cross section ratio.}
 \label{syserror}
 \end{center}
\end{table}

%%%%%%%%%%%%%%%%%%%%%%%%%%%%%%%%%%%%%%%%%%%%%%%%%%%%%%%%%%%%%%%%%%%%%%%%
%\section{Cross section ratio and Discussion}
%%%%%%%%%%%%%%%%%%%%%%%%%%%%%%%%%%%%%%%%%%%%%%%%%%%%%%%%%%%%%%%%%%%%%%%%
  Our result is consistent with the non-existence of CC coherent
pion production at K2K neutrino beam energies, and hence we set an upper 
limit on the cross section ratio at 90\% C.L. :
\begin{eqnarray*}
  \sigma(\mbox{CC coherent $\pi$})/\sigma(\nu_{\mu}\mathrm{CC}) < 0.60 \times
  10^{-2}.
\end{eqnarray*}
 For reference, the total CC cross section is
calculated as $1.07 \times 10^{-38}\mathrm{cm}^{2}/\mathrm{nucleon}$
in the neutrino MC simulation by averaging over K2K neutrino beam
energies. 

 The obtained upper limit is inconsistent with the model prediction 
by Rein and Sehgal at the level of 2.5 standard deviations.
We assign a 35~\% uncertainty to the theoretical 
prediction as described in~\cite{Rein:1982pf}.
 In addition, a finite cross section was reported by Aachen-Padova group
 for NC coherent pion production with 2~GeV average neutrino energy
 and with aluminum target~\cite{Faissner:1983}.
 If we assume an $A^{1/3}$ dependence of 
 the cross section ($\sigma$) and $\sigma(\mathrm{CC}) =
 2\sigma(\mathrm{NC})$ 
 according to the model of Rein and Sehgal, 
 the discrepancy between the extrapolation from 
 the NC measurement and the present result
 is as large as 3 standard deviations.
 There are other models predicting lower cross
 sections~\cite{Paschos:2003hs, Belkov:1986hn, Kelkar:1996iv}, 
but they do not provide the kinematics of pions and it is difficult
to test them directly. 
 Further theoretical work is necessary to construct 
interaction models which explain these experimental results.
 The non-existence of CC coherent pion production has given a solution to
the low-$q^{2}$ discrepancy observed in K2K.
It also reduces the uncertainty  
on the cross section in the relevant $q^{2}$ region, which is crucial
for the future neutrino oscillation experiments.

%%%%%%%%%%%%%%%%%%%%%%%%%%%%%%%%%%%%%%%%%%%%%%%
%\section{conclusion}
%%%%%%%%%%%%%%%%%%%%%%%%%%%%%%%%%%%%%%%%%%%%%%%

 In summary, we report on a search for CC coherent pion 
production by muon neutrinos with a mean energy of 1.3~GeV.
 The data analyzed correspond to $1.7 \times 10^{19}$ POT recorded 
with the K2K-SciBar detector.
 No evidence of CC coherent pion production is found and an upper limit 
on the cross section ratio of CC coherent pion production to the total
CC interaction is derived to be $0.60 \times 10^{-2}$ at 90\%~C.L.
 This result is the first experimental limit for CC coherent pion 
production by neutrinos with energies of a few GeV. 

%\section{Acknowledgements}
\begin{acknowledgments}
  We thank the KEK and ICRR directorates for their strong support
  and encouragement.
  K2K is made possible by the inventiveness and the
  diligent efforts of the KEK-PS machine group and beam channel group.
  We gratefully acknowledge the cooperation of the Kamioka Mining and
  Smelting Company.  This work has been supported by the Ministry of
  Education, Culture, Sports, Science and Technology of the Government
  of
  Japan,
  the Japan Society for Promotion of Science, the U.S. Department
  of Energy,
  the Korea Research Foundation,
  the Korea Science and Engineering Foundation,
  NSERC Canada and Canada Foundation for Innovation,
  the Istituto Nazionale di Fisica Nucleare (Italy),
  the Spanish Ministry of Science and Technology,
  and Polish KBN grants: 1P03B08227 and 1P03B03826.
\end{acknowledgments}

\end{document}